\documentclass[preprint,showpacs,preprintnumbers,amsmath,amssymb]{revtex4}

\usepackage{graphicx}

\begin{document}


\title{Sensitive Detection of Cold Cesium Molecules by Radiative Feshbach Spectroscopy}

\author{Cheng Chin, Andrew J. Kerman, Vladan Vuleti\'{c}, and Steven Chu}
\affiliation{Physics Department, Stanford University, Stanford, CA 94305-4060, USA}

\date{\today}
           
\begin{abstract}
We observe the dynamic formation of $Cs_2$ molecules near Feshbach resonances in a cold sample of atomic cesium using an external probe beam. This method is $300$ times more sensitive than previous atomic collision rate methods, and allows us to detect more than 20 weakly-coupled molecular states, with collisional formation cross sections as small as $\sigma =3\times 10^{-16}$cm$^2$. We propose a model to describe the atom-molecule coupling, and estimate that more than $2 \times 10^5$ $Cs_2$ molecules coexist in dynamical equilibrium with $10^8$ $Cs$ atoms in our trap for several seconds.
\end{abstract}

                             
                           
\maketitle
The newly emerging field of cold molecules provides intriguing possibilities in ultracold gas studies. In the realm of ``super-chemistry", phase-coherent chemical reactions between particles in well-defined internal and external states are dominated by quantum statistics and many-body effects \cite{superchem}. The coherent nature of the atom-molecule coupling has been recently observed in Bose-Einstein condensates \cite{becosci}. Quantum manipulation of molecular states has also been proposed for quantum computation \cite{qcom}. Finally, in precision tests of time-reversal symmetry based on matter-wave interferometry, polar molecules promise a sensitivity several orders of magnitude higher than atoms \cite{edmmolecule}.

However, the production of cold molecules presents a major experimental challenge. Direct laser cooling of molecules has been proposed using either multiple laser frequencies \cite{lasercoolm} or an external cavity \cite{cavity}; however, these methods involve exquisite experimental control, and have not yet been implemented. Photoassociation of cold atoms into molecules is limited by the phase space density of the initial atomic sample \cite{pa}, and consequently only a few thousand molecules have been produced from thermal samples with this method. Two-photon photoassociation from an atomic Bose-Einstein condensate has been limited to similarly small numbers of molecules due to large atom-molecule inelastic collision losses \cite{bec2photon}. Alternative routes to cold molecules by slowing down a supersonic molecular jet \cite{jet}, or by buffer gas loading of a magnetic trap \cite{doyle}, can collect large numbers of molecules, but the resulting temperatures are much higher than in the approaches using cold atoms.

In this work, we employ a radiation-induced loss mechanism to monitor the dynamic formation of cold $Cs_2$ molecules near Feshbach resonances. Our method is similar to the detection technique introduced by Heinzen and coworkers \cite{heinzen98}; however, instead of photoassociating ground state atom pairs, which requires knowledge of the excited-state molecular level structure, we use a blue-detuned probe beam to photodissociate them. In a dense sample of optically trapped cesium atoms we observe a class of weakly-coupled $Cs_2$ molecular ground states which, in the absence of the probe light, are collisionally metastable at the density of our samples. Since these states are long-lived, a large molecular population can build up even in a thermal sample.

\begin{figure}
\includegraphics[width=3.1in]{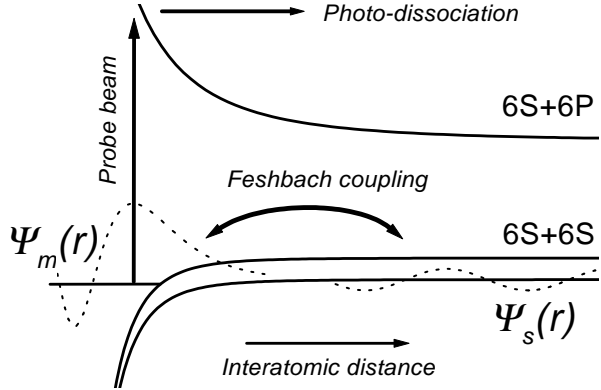}
\caption{Illustration of radiative Feshbach spectroscopy. In the vicinity of a Feshbach resonance the scattering state $\Psi_{s}$ couples to a ground molecular state $\Psi_{m}$. A probe beam tuned to the blue of the free-atom transition excites the molecules, which then dissociate and are lost from the trap due to the energy imparted by the strongly repulsive excited-state potential.}
\label{fig1}
\end{figure}

Our method for detecting the formation of ground state molecules, shown in Fig.~\ref{fig1}, can be understood as a radiative collision process \cite{bur96, vul99}. For a typical detuning of $\Delta/2\pi \sim +1$ THz from the free-atom transition, radiative loss occurs when the dipole-dipole interaction $V_{d}(r)\approx C_3/r^3$ between two colliding atoms shifts the excited state into resonance with the laser field at the Condon point $r_c$, where $r_c \approx 2$nm satisfies $\hbar\Delta=V_d(r_c)$ \cite{bur96}. The radiative coupling amplitude is therefore proportional to the two-atom wavefunction at the Condon point, $\Psi (r_c)$. On a Feshbach resonance \cite{tie9293}, the scattering state $\Psi_s(r)$ couples to a molecular state $\Psi_m(r)$, that has a much larger amplitude at short range and consequently at the Condon point. The coupling amplitude is then enhanced by a factor of $A\sim|\Psi_m(r_c)|/|\Psi_s(r_c)|\sim  (r_a/r_m)^{3/2}$, where $r_a$ ($r_m$) is the mean interatomic separation in the scattering (molecular) state. Typically, $r_a\sim 1\mu$m at our atomic density of $10^{13}$ cm$^{-3}$. $r_m$ can be estimated from the ground state van der Waals interaction $V_g=-C_6/r^6$, and the binding energy of the molecular state, which is on the order of the atomic hyperfine splitting $H_h$ \cite{tie9293}. Given $C_6=6859$ a.u. \cite{leo00} and $H_h/h= 10GHz$, we find $r_m \approx (C_6/H_h)^{1/6} \sim 2$nm. The radiative coupling amplitude of the molecule by the far blue-detuned beam is therefore a factor of $A\sim 10^{4}$ stronger than that of two trapped atoms. This indicates that molecules can be detected efficiently without appreciably perturbing the atoms.

\begin{figure}
\includegraphics[width=3.5in]{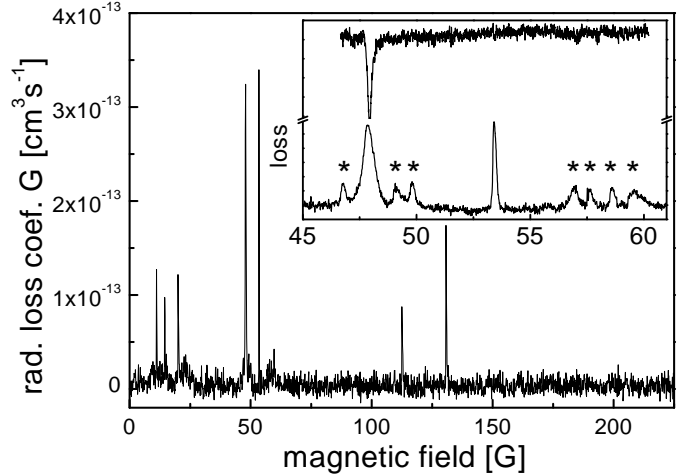}
\caption{Radiative loss coefficient as a function of magnetic field for a mean density of $1.5\times 10^{13}$cm$^{-3}$, $95\%$ population in $|F=3,m_F=3\rangle$, and a temperature of $5.5\mu$K. The wavelength and the mean intensity of the probe beam are $846.52$nm and $33$W/cm$^2$, respectively. In the inset, the radiative spectrum (lower curve) shows higher sensitivity than the collision rate measurement (upper curve). The stars indicate $|3,3\rangle +|3,2\rangle$ collision resonances.} 
\label{fig2}
\end{figure}

Compared to lighter alkalis, cesium atoms display unusually rich cold-collision phenomena, due primarily to the large dipolar interaction \cite{ju}. The angular dependence of this interaction allows an incident partial wave with angular momentum $L$ to couple to molecular states with orbital angular momentum $N=L\pm 2n \hbar$, in $nth$ order \cite{tie9293}. Consequently, many molecular states become available to scattering atom pairs when this interaction is strong. In the following, the above process is referred to as an $L\rightarrow N$ resonance, e.g. an $S\rightarrow G$ resonance occurs when an s-wave scattering state couples to a molecular state with an orbital angular momentum $N=4\hbar$.

In this work, we apply radiative Feshbach spectroscopy to two types of cold collisions: When both scattering atoms are in the lowest hyperfine ground state $|F=3,m_F=3\rangle$, denoted in the following by $|3,3\rangle +|3,3\rangle$, the scattering state $\Psi_s$ can couple to molecular states $\Psi_m$ with even orbital angular momentum as the relative energy of $\Psi_s$ and $\Psi_m$ is tuned by means of a magnetic field. We find multiple $S\rightarrow D$, $S\rightarrow G$ Feshbach resonances, and one $S\rightarrow D$ shape resonance for samples polarized in $|3,3\rangle$ \cite{fromeite}. For $|3,3\rangle +|3,2\rangle$ collisions, on the other hand, both even  and odd partial waves are allowed, and we observe an abundance of $S\rightarrow D$, $S\rightarrow G$, $P\rightarrow P$ and $P\rightarrow F$ Feshbach resonances \cite{fromeite}.

Our experimental setup has been described previously in Ref.~\cite{vul98}. About $3\times 10^8$ atoms are loaded into a far-detuned dipole trap formed by a YAG laser at $1064$nm using two pulses of Raman-sideband cooling \cite{vul98, ker00}. We then further optically pump so that up to $98\%$ of the remaining atoms are in $|3,3\rangle$, and by adjusting the optical pumping frequency prepare mixed samples as desired with up to $10\%$ in $|3,2\rangle$. For all of the experiments described here, less than $1\%$ ($0.25\%$) of the atoms are found to occupy $|3,1\rangle$ (all other states), as has been determined by microwave spectroscopy \cite{chi}. Within $15$ms the atoms  thermalize to a temperature between 3$\mu$K and 6$\mu$K, and density near $10^{13} cm^{-3}$ by means of elastic collisions; the atom number is measured by fluorescence detection and the density is then calculated from the measured temperatures and trap vibration frequencies. We adiabatically ramp the magnetic field from $0.1$G to $2$G in $1$ms, and then to any desired field value up to $230$G in another $1$ms. No depolarization of the sample is observed during this procedure. A Titanium-Sapphire laser provides the far-detuned probe beam with an $e^{-2}$ beam waist of $2.0$mm ($0.6$mm) in the vertical (horizonal) direction, which uniformly illuminates the sample whose vertical (horizontal) size is $580\mu$m ($60\mu$m) \cite{vul99}. The typical mean intensity and wavelength, optimized for radiative detection sensitivity, are $20$W/cm$^2$ and $847$nm, respectively.  Loss coefficients are extracted from the evolution of the atomic density, assuming a Gaussian distribution, according to $\dot{n}=-n/T-Gn^2-(4/3)^{3/2}Kn^3$, where $n$ is the mean atomic density, $T=3$ s is the one-body lifetime, and $G(K)$ is the two (three)-body decay coefficient.

\begin{figure}
\includegraphics[width=3.5in]{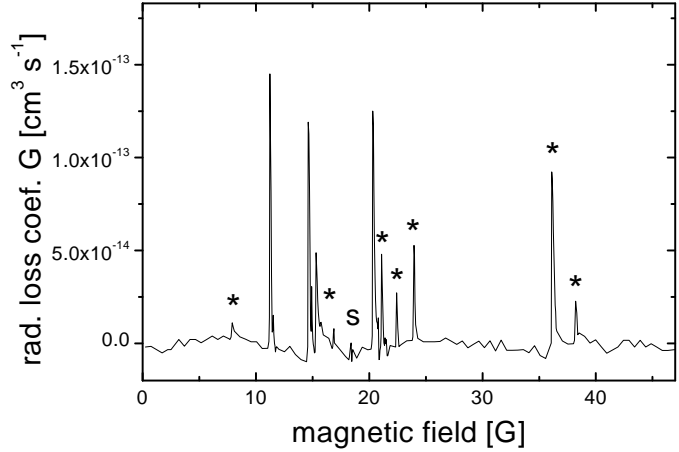}
\caption{A detailed radiative loss spectrum with a resolution of $5$mG for a probe beam at $844$nm. $90\%$ ($10\%$) of the atoms are in the $|3,3\rangle$ ($|3,2\rangle$) state. The temperature and density of the sample are $3.5\mu K$ and $1.8\times 10^{13} cm^{-3}$, respectively. ``S" indicates a shape resonance; the stars indicate $|3,3\rangle +|3,2\rangle$ collision resonances.} \label{fig3}
\end{figure}

Fig.~\ref{fig2} shows the radiative loss rate from $0$G to $230$G, measured with a resolution of $100$mG. All seven observed strong resonances are due to $|3,3\rangle+|3,3\rangle$ collisions. The lower curve in the inset shows a higher-resolution scan performed on a sample with a larger impurity in $|3,2\rangle$. Seven additional smaller resonances are evident, which originate from $|3,3\rangle+|3,2\rangle$ collisions. The upper curve in the inset is the corresponding measurement of elastic collision-induced evaporative loss performed under similar conditions \cite{chi}. 

It is worth noting that only the strongest resonance at $48.0$ G could be detected with the collision rate measurements in Ref.~\cite{chi}. There are two reasons for this: first, there is no appreciable trap loss due to inelastic collisions near these resonances. In particular, for $|3,3\rangle+|3,3\rangle$ collisions, where binary inelastic collisions are endothermic and therefore forbidden, the resonant three-body loss is observed to be below our detection resolution of $K_3 = 3\times 10^{-27}$ cm$^6$/s even on the strongest $S\rightarrow G$ wave resonance at $53$ G. Second, these resonances are sufficiently narrow in energy that the thermally averaged change in elastic collision rate is smaller than our detection limit of $3\%$. Clearly the radiative Feshbach resonance detection offers higher sensitivity and accuracy for locating molecular states.

The inset to Fig.~\ref{fig2} shows two complementary aspects of Feshbach resonances: When the field shifts a molecular state into resonance with the scattering continuum, an increase in radiative loss indicates the appearance of a molecular population; at the same time a change in evaporation loss rate indicates a resonant alteration of the effective elastic collision cross section. The difference between these two effects is exemplified by the fact that the radiative resonance at $47.97$ G occurs at a slightly lower field value than does the minimum in the evaporation loss at $48.02$ G. While the radiative resonance occurs when the molecular state is tuned to the scattering continuum (Feshbach resonance), the minimum in the elastic cross-section occurs when the background scattering length is cancelled by the resonant contribution from the bound state (Ramsauer-Townsend effect). The finite width of the resonance causes these two points to be distinct. For stronger (wider) resonances this is more dramatic: the wide minimum in the elastic cross section observed at $17.1$G in Ref.\cite{chi} is in fact induced by the Feshbach resonance at a negative field of $-8.25$G (or equivalently, positive field for the $|3,-3\rangle$ state) \cite{fromeite}.

We model the radiative process using a chemical rate-equation approach that treats atoms and molecules as two different species \cite{timmermans}. In a thermal sample of de Broglie wavelength $\lambda_{dB}$ with low phase space density $\phi = n \lambda_{dB}^3 \ll 1$, the evolution of the mean atomic (molecular) density $n$ ($m$) near a Feshbach resonance can be described classically: Based on detailed balance between one molecular state and $1/\phi$ scattering states in the relative coordinate, $m=\frac{1}{2} \phi n$ molecules per unit volume can coexist with $n$ atoms in thermal equilibrium in the absence of collision loss \cite{nphi}. The formation and destruction rates of molecules are then described by $\alpha \phi n/2$ and $-\alpha m$, where $\alpha$ characterizes the strength of the Feshbach resonance. Given a leading order collisional loss $-\gamma n m$ due to atom-molecule collisions, and radiative atom loss induced by the probe beam $L=2 \delta m I$, where $\gamma$ and $\delta$ are rate constants, and $I$ is the probe beam intensity, we obtain the following rate equations:

\begin{eqnarray}
\dot{m}&=&\alpha\, \phi n/2-\alpha\, m-\gamma\, n m - L/2
\label{eq1} \\
L&=&2 \delta\, I m
\label{eq3}.
\end{eqnarray}

The stationary solution for the molecular density $m$ and the fractional atom loss rate $L/n$ are then given by

\begin{eqnarray}
m&=&\frac{(\phi/2) n}{1+(\gamma/\alpha) n + (\delta/\alpha) I} \label{eq4}\\
L/n&=&\frac{\delta I \phi}{1+(\gamma/\alpha) n + (\delta/\alpha) I}. \label{eq5}
\end{eqnarray}

\begin{figure}
\includegraphics[width=3.3in]{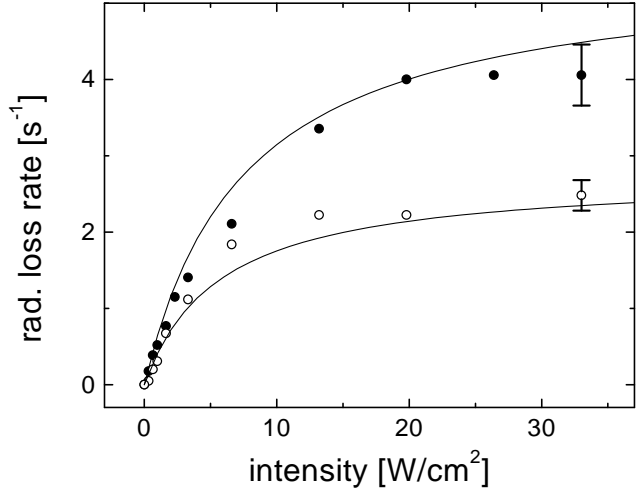}
\caption{Radiative loss rate at $53.5G$ vs. intensity for atomic densities of $n_1=1.2\times 10^{13}$cm$^{-3}$(solid circles) and $n_2=0.6\times 10^{13}$cm$^{-3}$(open circles). The wavelength of the probe beam is $846.5$nm, and the temperature of the sample is $5.5\mu$K. We fit the two curves with Eq.~\ref{eq5}.}
\label{fig4}
\end{figure}

To test this model, we measure the radiative atom loss as a function of probe beam intensity $I$ for different atomic densities, as shown in Fig.~\ref{fig4} for the strongest $S\rightarrow G$ resonance at $53.50$ G. Here the mean atomic density is switched between $n_1=1.2\times 10^{13}$cm$^{-3}$ and $n_2=0.6\times 10^{13}$cm$^{-3}$ by changing the trap loading parameters. The phase space densities are $\phi_1 \approx 1/100$ and $\phi_2 \approx 1/200$, respectively. The loss constant $L/n$ is derived from the measured radiation exposure time $t$ necessary to deplete $10\%$ of the sample, namely, $L/n=0.1/t$. The polarization purity of $95\%$ in $|3,3\rangle$ and the temperature of $5.5\mu$K do not vary systematically by more than $5\%$ between the two settings. We have verified that no additional loss is due to field ramping or radiative processes induced by the very far detuned dipole trap at $1064$nm. 

At low probe beam intensity $I<5$W/cm$^2$, the fractional loss rate is linear in $I$. This suggests that the molecular and atomic populations reach a dynamic equilibrium where the probe beam only weakly depletes the molecules, without appreciably changing their density. At high probe beam intensity $I>10$W/cm$^2$, the fractional loss rate is  proportional to the sample density and is independent of the probe beam intensity. This implies a quick depletion of molecular population by the probe beam, leading to a loss rate that is therefore equal to the formation rate of the molecules. As indicated by the solid lines in Fig.~\ref{fig4}, Eq.~\ref{eq5} agrees well with the experimental results, and a simultaneous fit to both sets of data with only two free parameters yields $\alpha =364(20) $s$^{-1}$ and $\delta/\gamma=3.3(6) \times 10^{12}$(Wcm)$^{-1}$

From the above measurements, we can first deduce the sensitivity of radiative Feshbach spectroscopy for detecting molecule formation. Given $\alpha$, the molecule formation cross section at $53$ G is $\sigma_{\alpha}=\alpha \phi/v n=8\times 10^{-14}$cm$^2$, where $v$ is the mean velocity of the atoms. While this resonance induces a maximum loss rate of $4$s$^{-1}$, the loss rate of the weakest observed resonance at $7.8$G is approximately $0.04$s$^{-1}$, with a signal-to-noise ratio of $3$. We estimate that the sensitivity of our detection, expressed in terms of a molecule formation cross section, is then $\sigma_{\alpha}/300 \approx 3\times 10^{-16}$cm$^2$. Compared to the nominal elastic cross section for atoms in $|3,3\rangle$ of $1.5\times 10^{-10}$cm$^2$, this sensitivity corresponds to the creation of less than two molecules every $10^6$ collisions. 

Furthermore, we can calculate the stationary number of molecules in the gas from the measured rate constants. While the latter only reveal the formation constant $\alpha$, we can safely put an upper limit on the collisional loss $\gamma$ from the unitarity limit of the s-wave cross section for inelastic collisions: $\sigma<\pi/k^2$. We then estimate $\gamma=\sigma_{\gamma}v_{\gamma}< (\pi /k^2) (\hbar k/\mu)$, where $\hbar^2 k^2/2\mu\sim \frac32 k_B T$ is the collision energy, $\mu=\frac23 M$ the reduced mass, and $M$ the mass of the Cs atom. Given Eq.~\ref{eq4} and $\alpha$, we obtain $\gamma<2.8\times 10^{-11}cm^{3}$s$^{-1}$ and a ratio of molecular to atomic density $m/n>2.6(5)\times 10^{-3}$ at the atomic phase space density of $\phi=1/100$ and low radiative excitation. Equivalently, more than $2.6(5)\times 10^5$ molecules coexist with $10^8$ atoms in our trap.

In conclusion, we have used radiative Feshbach spectroscopy to detect more than twenty new $Cs_2$ molecular states with high sensitivity and resolution. Since no strong collision loss accompanies these resonances, as many as $3 \times 10^5$ molecules accumulate and coexist dynamically with $10^8$ atoms in our dipole trap. In the future, we hope to explore the regime where the atom-molecule coupling strength exceeds the thermal energy of the gas, and its coherent nature may become evident even in our thermal sample.

This work was supported in parts by grants from AFOSR and the NSF. 

\end{document}